\pdfoutput=1

\documentclass[preprint,12pt]{elsarticle}

\usepackage[T1]{fontenc}
\usepackage{lmodern}
\usepackage[utf8]{inputenc}
\usepackage{textcomp}
\usepackage{amsmath,amssymb}
\usepackage{graphicx}
\usepackage{booktabs}
\usepackage{array}
\usepackage{tabularx}
\usepackage{newunicodechar}
\usepackage{xcolor}
\usepackage[hidelinks]{hyperref}

\newcolumntype{L}{>{\raggedright\arraybackslash}X}
\newcolumntype{C}{>{\centering\arraybackslash}X}
\newunicodechar{°}{\textdegree}
\newunicodechar{−}{\textminus}
\newunicodechar{κ}{\ensuremath{\kappa}}
\newunicodechar{×}{\ensuremath{\times}}
\newunicodechar{³}{\textsuperscript{3}}
\makeatletter
\def\ps@pprintTitle{%
  \let\@oddhead\@empty\let\@evenhead\@empty
  \def\@oddfoot{\footnotesize\itshape Preprint\hfill}\let\@evenfoot\@oddfoot}
\makeatother

\begin{document}

\begin{frontmatter}

\title{Narrow and Deep: An Ontology Tower as the Knowledge of an LLM Agent for an Industrial Equipment System}

\author[kier,ust]{Younghwan Joo\corref{cor1}}
\ead{yhjoo@kier.re.kr}
\author[kier,ust]{Sung-il Kim}

\affiliation[kier]{organization={Energy Efficiency Research Division, Korea Institute of Energy Research},
  addressline={152 Gajeong-ro, Yuseong-gu},
  city={Daejeon},
  postcode={34129},
  country={Republic of Korea}}
\affiliation[ust]{organization={Energy Engineering, University of Science \& Technology},
  addressline={217 Gajeong-ro, Yuseong-gu},
  city={Daejeon},
  postcode={34113},
  country={Republic of Korea}}

\cortext[cor1]{Corresponding author}

\begin{abstract}
Large language model (LLM) agents are beginning to operate industrial energy equipment, and what they get right depends on what they are told about the plant. Established building ontologies name many kinds of points across many sites, whereas an industrial equipment system needs few entities with much knowledge about each. This study proposes the ontology tower, a narrow-and-deep ontology of a single equipment system whose knowledge deepens in two ways: through quantities derived from the measured points by physical relations, and through lessons from the operating journal incorporated as knowledge nodes. On a real low-humidity air-handling test plant operated daily through a programmable logic controller, agents received a text projected from its tower in a preregistered evaluation of nine tasks replayed from the plant's records, using four open-weight models from 9 to about 750 billion parameters. This knowledge raised the rate at which the agents avoided the most plausible misjudgment of each task by about 20 percentage points, and the overall task score of the 9-billion-parameter model as much as that of the largest. Operating lessons were used when incorporated into the tower or placed in the prompt as records, but seldom when left in the journal behind a search tool. In live runs through an invariant safety layer, the agents brought the controlled variable into its target band in 12 of 14 runs. An ontology narrow in entities but deep in what is known about them can thus supply the knowledge that an agent for an industrial equipment system needs.

\end{abstract}


\end{frontmatter}

\section{Introduction}
\label{sec:1}

Large language model (LLM) agents are beginning to take part in operating building and industrial energy equipment: they read process points, diagnose faults, and propose setpoint changes~\cite{yoon.etal2025OntologyEnabled,chen.etal2025Customized,coLLM2026Chiller,jarvis2025HVACQA}. Because plant data are often kept within the operator's own network, such agents are also being built on open-weight models that run on site~\cite{openweight2026WindAgent,lightweight2026BEMS}. In these studies the agent receives one package of knowledge, in which the meaning of the tags, the couplings of the equipment, the procedures, and past incidents enter together, and the package is judged as a whole against a baseline without it~\cite{zhang.etal2025HVACFineTuning,assetopsbench2026}. However, what the knowledge of a particular plant should contain, and in what form it should reach the agent, has not been measured. Therefore, a structure is needed in which the knowledge of one plant is kept explicitly and can be supplied to an agent in controlled parts.

Ontologies are the established way to keep such knowledge. Brick~\cite{balaji.etal2018BrickMetadata}, SAREF~\cite{daniele.etal2015SAREF}, the Asset Administration Shell~\cite{iec2023AAS}, and ISO 15926~\cite{iso2004ISO15926} define vocabularies that name points and assets across many buildings and plants, so that applications can find them. Agents that query such models through graph retrieval have answered building questions~\cite{buildinggpt2025}, supported multi-zone control~\cite{bhatt.etal2026ThermoLLM}, and outperformed agents without a graph-based data layer~\cite{mandarapu.kunkunuru2026AssetKG}. These vocabularies are broad by design: they cover many kinds of equipment with a shallow description of each. An agent that operates one equipment system needs the opposite. The system has few entities, but the agent must know how each value is derived from the measured points, which assumptions it depends on, and what has happened to the system in operation. However, no ontology for equipment states the physical derivations and the operating history of a single system together with its points. Therefore, an ontology that is narrow in its entities and deep in its knowledge is required.

How much an agent gains from supplied knowledge may also depend on the model, and the reported results disagree. Retrieval improved smaller models the most in Ren et al.~\cite{ren.etal2023Knowledge}, and a 14-billion-parameter model with retrieval approached GPT-4 in clinical decision support~\cite{park.etal2026RAGClinical}. In contrast, Ning et al.~\cite{ning.etal2026LessLLM} found the largest gain at intermediate sizes, and Pandey~\cite{pandey.etal2026SmallLMs} found that models of up to seven billion parameters failed to use retrieved passages even when the passages held the answer. Luong Tuan and Sanyal~\cite{tuan.sanyal2026Ontology} attributed the effect of ontology grounding to how well training covered the domain rather than to model size. None of these studies supplied curated knowledge of equipment or concerned a physical plant. Therefore, whether a small open-weight model gains as much as the largest open-weight models from the knowledge of a plant, and whether the form of supply matters, has to be measured on real equipment.

A plant also changes while it is operated, and what its operators learn is first written into an operating journal of incidents, investigations, and corrections. Systems that build a knowledge base from operating logs retrieve from it as a step of their procedure~\cite{coLLM2026Chiller}, and the maintenance of an ontology whose content changes over time has been studied as ontology evolution~\cite{stojanovic.etal2002OntologyEvolution,noy.klein2004OntologyEvolution}. However, whether an agent uses the lessons of operation when they are incorporated into the ontology, compared with when they are left in the journal for the agent to retrieve, has not been tested. Therefore, an ontology for an agent should state how it takes in the lessons of operation, and that choice should be tested within the same tasks.

In this study, an ontology tower, a narrow-and-deep ontology of a single equipment system, is proposed as the knowledge that an LLM agent receives about a plant, and it is tested on a real low-humidity air-handling test plant operated daily through a programmable logic controller (PLC). To give the ontology depth, quantities derived from the measured points by physical relations are added above the tagged points, with the level of each node computed from its declared dependencies, and lessons from the operating journal are attached to the points they concern as knowledge nodes. To supply it to an agent, a text projected from the tower is placed in the prompt, and every command that the agent issues passes through an invariant safety layer. To evaluate this supply, a preregistered evaluation of reading and diagnosis tasks replayed from the plant's records was conducted with four open-weight models, together with comparisons of the form in which the knowledge and the operating lessons are supplied. Finally, the agents changed setpoints on the live plant through the safety layer. The evaluation is confined to one plant, and its extension to many similar systems is discussed in Section~\ref{sec:4}.

\section{Method}
\label{sec:2}

In this section, the ontology tower is presented first: the principle of narrow and deep knowledge, the tower and its height, and the two ways in which its knowledge deepens. The supply of the tower to an agent and the safety layer that bounds the agent's actions are then described. Finally, the test plant and the three tests are presented: a preregistered evaluation, an experiment on the form of the operating lessons, and live runs on the plant.

\subsection{Narrow and deep knowledge of one equipment system}
\label{sec:2.1}

An ontology tower covers one equipment system, that is, several machines operated as one unit, such as an air handler with its chiller and desiccant rotor. The number of entities is small, a few dozen measured points and setpoints, but each is described far beyond its name: its binding to the controller, its unit and scale, how it was verified, which quantities are derived from it, which assumptions those derivations depend on, and what the operators have learned about it. Broad vocabularies such as Brick name the points that such a system exposes; the tower stands on those points and holds what an agent needs to reason about this one system.

An ontology tower (OntoTower) is thus a narrow-and-deep ontology of a single equipment system: a set of physically grounded tags, assumptions treated as nodes, derived quantities, and knowledge nodes, in which the level of each node is computed from the dependencies declared for it rather than assigned by the modeler.

\subsection{The tower and its height}
\label{sec:2.2}

\begin{figure}[!htbp]
\centering
\includegraphics[width=\textwidth]{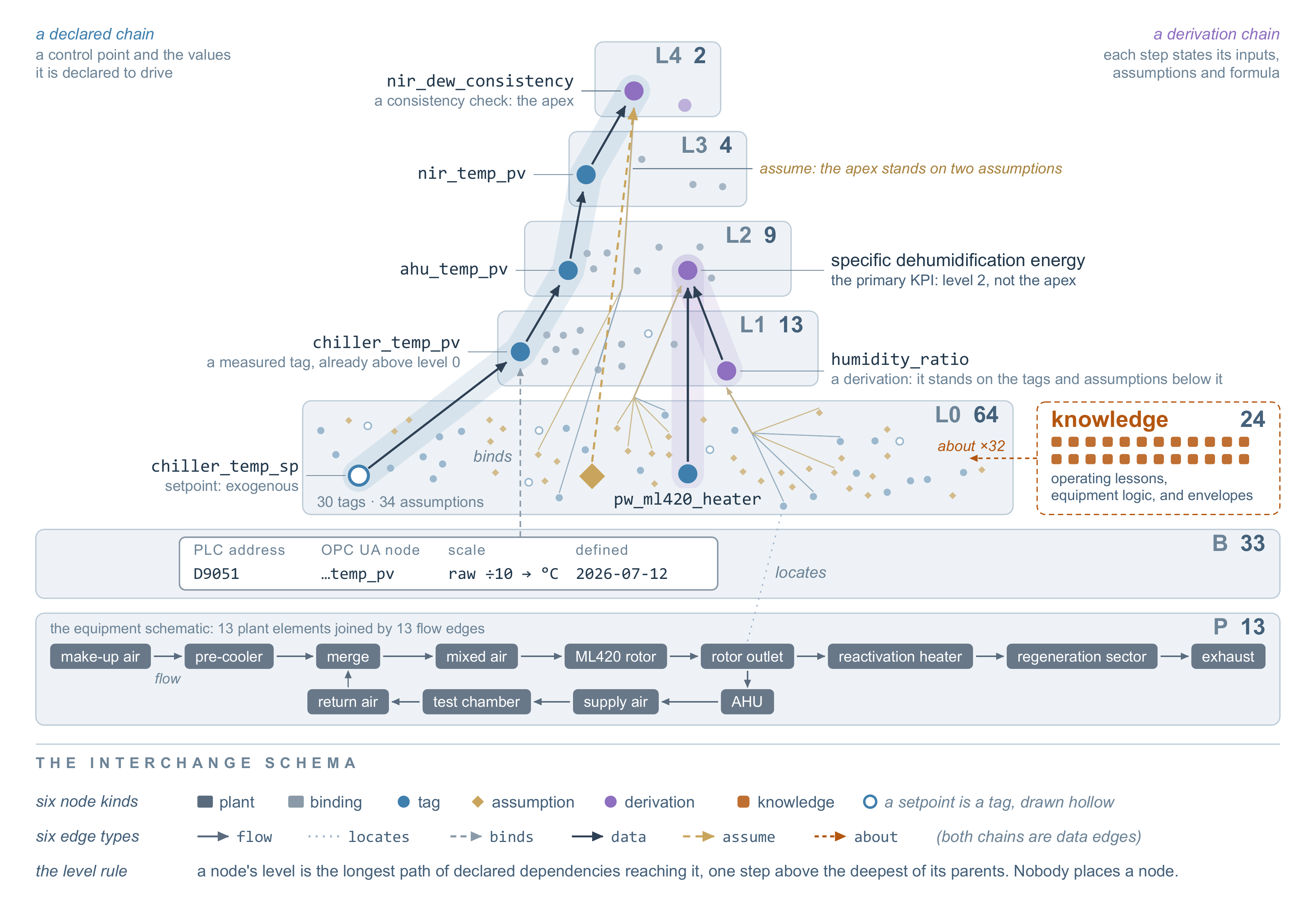}
\caption{Ontology tower of the studied system: node counts at each computed level, two dependency chains that fix the levels, and the node and edge kinds of the interchange schema.}
\label{fig:1}
\end{figure}

Fig.~\ref{fig:1} shows the tower of the studied system, drawn from its own data. The height of the tower shows how far its knowledge extends above measurement. The bottom stratum, P, is the equipment schematic. The stratum above it, B, binds each tag read from the controller, a named process point of the control system such as a supply-air temperature or a setpoint, to its PLC address, raw scale, unit, and the code that defines it; in the studied tower, 33 of the 42 tags have such a binding. Level 0 holds the tags, the assumptions, and the knowledge nodes, which Fig.~\ref{fig:1} draws beside the tower, and every level above it holds what depends on the levels below.

The height is computed from declared dependencies; it is neither drawn by hand nor read off the physics. The level of a node is the length of the longest path of dependencies that reaches it from measurement, one step above the deepest of its parents. In a real plant nearly every quantity influences many others, so a dependency edge is drawn only where the plant's own repository declares a directed relation, and there are three such sources: the inputs of a derivation, a control point and the process variables it is declared to drive, and a causal chain recorded among the couplings of the units. Constraint and coordination relations, such as an upper limit that one controller places on another or the split of airflow between branches, are kept as knowledge and do not become edges, and relations without a natural direction, such as conservation laws, become checks that depend on both sides. A tag therefore rises above level 0 only when such a declaration names it as driven. In the studied tower, 12 of the 42 tags do; the highest is the chamber temperature at level 3, reached through the declared chain from the chilled-water setpoint through the supply-water and supply-air temperatures (Fig.~\ref{fig:1}, left). A level thus states how far a value lies from measurement in the declared knowledge, not in the physical field.

The computed levels need not agree with how a person perceives the system. A person looking at this system sees three strata, raw points, first derivations, and performance indicators, whereas the computation gives five levels, places the main performance indicator, the specific dehumidification energy, at level 2, and places two consistency checks at the top. A level states how far a value lies from measurement, not what the value is used for. The tower of this system has 162 nodes, namely 42 tags, 34 assumptions, 33 bindings, 24 knowledge nodes, 16 derived quantities, and 13 plant elements, joined by 217 edges.

\subsection{Depth through physics}
\label{sec:2.3}

The first way in which the knowledge deepens is through physics, and it is recorded in the height of the tower. Above the measured points, the tower holds quantities derived through engineering relations, such as psychrometric conversions, mass and energy balances, and fan affinity laws, and each derivation states its inputs, its formula, its unit, and the assumptions it uses. The 16 derived quantities of the studied tower occupy levels 1 to 4, including the two consistency checks at the top. In the right chain of Fig.~\ref{fig:1}, the power of the reactivation heater, a measured airflow, and a humidity ratio combine into the specific dehumidification energy. An assumption is anything believed but not measured, such as an atmospheric pressure taken as 1013 hPa or a flow sensor trusted while its re-verification is pending, and it is a node with a statement and a status that changes when the belief is confirmed by measurement. Because every derivation is linked to the assumptions it uses, the tower shows which values rest on each belief: the assumed atmospheric pressure enters four derived quantities, among them the humidity ratio and the specific dehumidification energy, and the two assumptions on the flow sensors enter six. When such an assumption is checked by measurement and found wrong, these are the values to be recomputed. As the description of a system grows, this depth can extend inside the components themselves, for example to the flow within the coil of a heat exchanger, and each such quantity enters the tower as a derivation with its own inputs and assumptions. The ontology thus holds the physics of the system as well as its names.

\subsection{Depth through operating history}
\label{sec:2.4}

The second way in which the knowledge deepens is through time. A system changes while it is operated, and its operators record incidents, investigations, and corrections in an operating journal. When an incident yields a lesson, for example that a status flag reporting normal operation can hide an abnormal cycle of the desiccant unit, the lesson is written into the tower as a knowledge node linked to the nodes it concerns. The knowledge node kind holds operating lessons, the automatic logic of the equipment, and operating envelopes as text. This depth does not add to the height of the tower: a knowledge node stays at level 0, and its links to the nodes it concerns are not counted in the levels, so it adds to what is known about a point rather than to how far the point lies from measurement. Whenever the tower is extracted, the journal entries written since the previous extraction are incorporated in this way, so the tower is an evolving tower in the sense of ontology evolution~\cite{stojanovic.etal2002OntologyEvolution,noy.klein2004OntologyEvolution}. A lesson incorporated in this way is specific to the system that produced it, and the tower becomes more specific to its system the longer the system is operated. In the studied plant, this practice preceded the tower: of the four incidents used as diagnosis tasks below, two had their lessons written into the plant's ontology files when their journal entries were written, and the other two were never incorporated.

\subsection{Supplying the tower to an agent}
\label{sec:2.5}

\begin{figure}[!htbp]
\centering
\includegraphics[width=\textwidth]{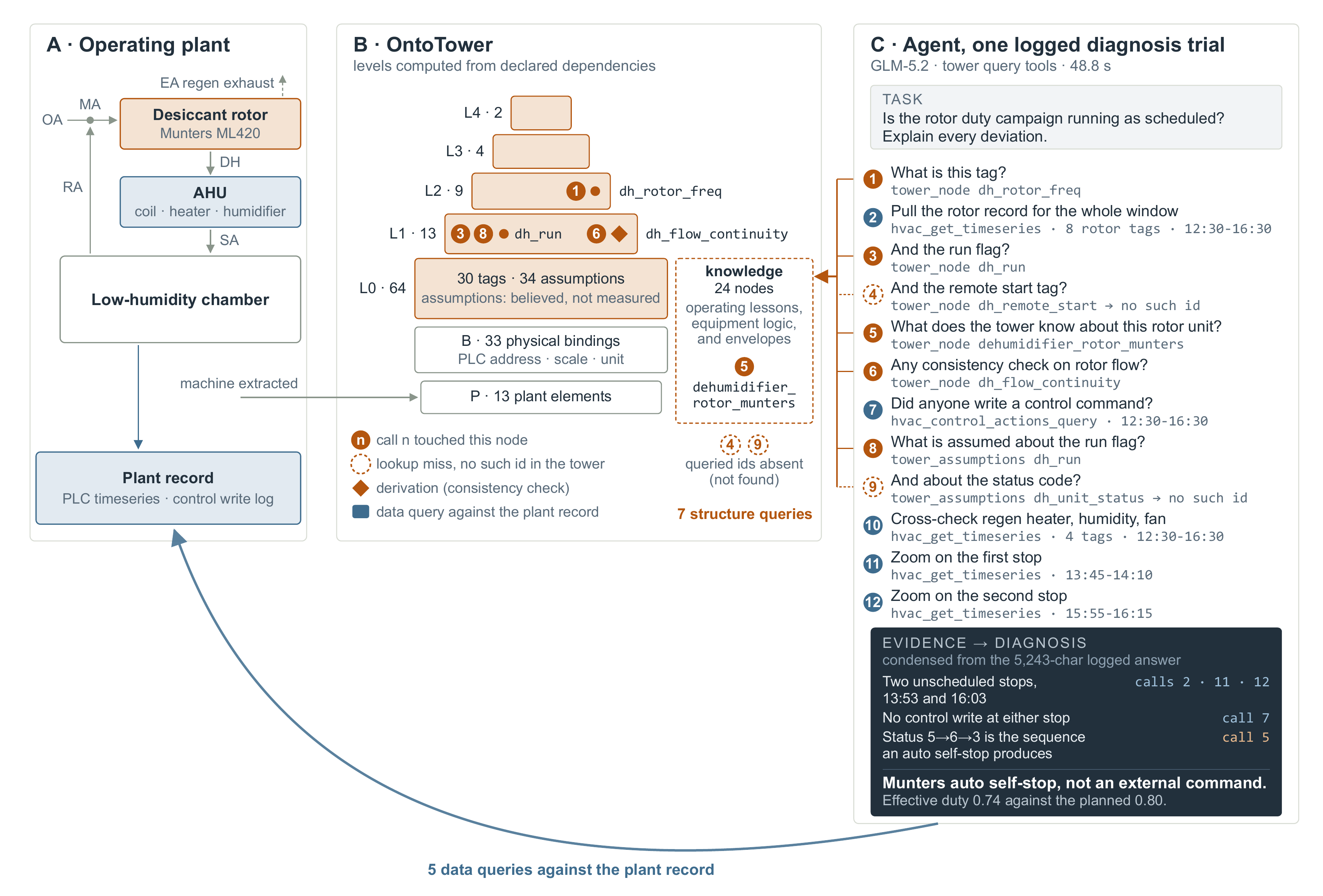}
\caption{One diagnosis trial with the tower query tools: (A) the plant, (B) the tower with the nodes queried, and (C) the sequence of calls and the diagnosis.}
\label{fig:2}
\end{figure}

The agent does not receive the nodes and edges of the tower; it reads a text projected from them. The projection states, in prose and structured blocks, the identity, verification, and known faults of each sensor, the couplings between the units and their automatic logic, the knowledge nodes, and the operating envelopes of the control points, in about 45,000 characters for this plant, and it is placed in the agent's prompt. The projection is generated from the same descriptive records from which the tower is extracted, namely the plant's sensor registry, the descriptions of the couplings, the ontology of its control points, and its system targets, so every entry of the projection appears in the tower. The plant keeps its derivations as computations in its monitoring software rather than as descriptions, and the 16 derived quantities and their assumptions are therefore in the tower but not in the projection, and the evaluation tests the tower's descriptive content (Section~\ref{sec:4.3}). The same knowledge can reach the agent in other forms, which the evaluation compares (Section~\ref{sec:2.8}): the whole tower serialized as JSON (about 99,000 characters), the tower itself offered to the agent as tools, namely three read-only queries with which the agent looks up a node, follows its edges, and lists its assumptions, and the plant's documents behind a search tool. With the query tools the agent consults the tower directly, so a statement that the agent bases on a query can be traced to the node that the query returned, a property that the text does not have. Fig.~\ref{fig:2} shows one diagnosis trial with the query tools, run on the plant's recorded data. Panel (A) shows the plant, panel (B) the tower with each node that the agent queried marked by the number of the call, and panel (C) the calls in order and a condensed diagnosis, beside whose findings the authors marked the calls that supplied the evidence. The agent made twelve calls, seven to the tower, two of which asked for nodes that do not exist, and five to the plant data.

\subsection{Bounding the agent's actions}
\label{sec:2.6}

An agent that may change setpoints acts through one write path. Every command passes through an invariant safety layer, the same code in every run, which checks in order whether the tag is on the list allowed for the run, whether the value lies within the permitted range, whether a person has written to the plant recently, and whether the change satisfies the guard that the plant's software applies to automated setpoint changes: absolute limits, step size, hourly and cumulative change limits, and a waiting time between changes. It then writes the value, reads it back, and records the command, accepted or refused.

\subsection{Test plant}
\label{sec:2.7}

The test plant is a low-humidity air-handling system of the kind used for industrial dry rooms. It was designed and built at the authors' institute with the equipment manufacturers to condition a test chamber, and several research groups use it daily. It is industrial in kind rather than a laboratory model: the desiccant unit, the air-handling unit, and the chiller are commercial equipment run by the plant's PLC~\cite{joo.etal2026AutomatedDigitalTwin}. Make-up air from the building's conditioned space is pre-cooled by a packaged unit, mixed with return air from the chamber, and dried in the process sector of a desiccant rotor (Munters ML420, rated at 420 m³/h of process air), whose outlet reaches a frost point near −40 °C. The dried air then splits in a ratio of about two to one. The larger stream passes the air-handling unit, which has a chilled-water coil served by a dedicated precision chiller, four 2.5 kW electric heaters, and an electric humidifier, and returns to the chamber as supply air. The smaller stream is heated electrically to about 130 °C and passed back through the regeneration sector of the rotor before it is exhausted. Dew-point transmitters measure the supply air, the rotor outlet, and the return air, and a fourth inside the chamber reads high and is not used for judgments; thermal mass-flow stations measure the principal air streams, and electric power is metered for each device. The plant's operating rules prefer the dew-point sensors to the relative humidity sensors, which are unreliable below about 5 \% relative humidity, the range in which this plant mostly operates. The operators change setpoints through the plant's supervisory control software or through the web interface of a server beside the plant, which logs every point at a nominal interval of 15 s and hosts the plant's LLM assistant.

\subsection{Preregistered evaluation}
\label{sec:2.8}

Nine tasks were built from the plant's records. In the five reading tasks, the agent receives a snapshot of the plant's display at a fixed time and must interpret specified points, for example a humidity point whose label suggests the wrong location. Each of the four diagnosis tasks is built on a real incident: an abnormal oscillation of the desiccant unit, a self-stop limit cycle of the same unit, a sweep of setpoint changes made from outside the agent's server, and a stop cascade at low airflow. Each task specifies ground-truth elements with points and trap elements, plausible misjudgments that the data invite. The element score of a trial is the share of the points awarded, which can slightly exceed 1 where a task has a bonus element, and the primary-trap avoidance rate is the share of trials that avoid the task's first and most plausible misjudgment. The agent worked in an evaluation copy of the plant's LLM assistant, limited to 60 steps and 1,500 s per trial, with read-only tools for points, values, time series, statistics, and the record of control actions, all cut at the time of the task, and with a general tool for reading files in its workspace. In these tasks the plant's records were replayed, and control of the live plant was tested separately (Section~\ref{sec:2.10}). Two diagnosis incidents that predate the retention of the time-series database were reconstructed from values in the operating journal.

The knowledge was supplied in five cumulative conditions, K0 to K4, and in six further conditions that remove one part of them or supply it in another form (Table~\ref{tab:1}). The comparison of K1 and K2 measures the projected text of the tower. K4x removes from K4 the records of the task's own incident, and the remaining five conditions replace a layer with another form of the same knowledge. Table~\ref{tab:1} also shows which knowledge of Sections~\ref{sec:2.2} to \ref{sec:2.4} each condition carries: the incorporated lessons reach the agent in K2 as knowledge nodes, the journal records only through the tools of K4, and the derived quantities only in the two tower conditions.

\begin{table}[!htbp]
\centering
\small
\caption{Knowledge conditions of the evaluation and the knowledge of the tower that each carries.}
\label{tab:1}
\begin{tabularx}{\textwidth}{>{\hsize=.76\hsize}L>{\hsize=1.12\hsize}L>{\hsize=1.12\hsize}L}
\toprule
\textbf{Condition} & \textbf{What the agent receives in addition} & \textbf{Knowledge of Sections~\ref{sec:2.2} to \ref{sec:2.4}} \\
\midrule
K0, address level & PLC addresses and raw integer values only & Bindings without names \\
K1, tag level & Tag names, units, and quality flags, as on the plant's display & Tags \\
K2, structured knowledge & The text projected from the tower (Section~\ref{sec:2.5}) & Tag attributes, couplings, envelopes, and knowledge nodes with the incorporated lessons; no derived quantities \\
K3, procedural & Summaries of the six operating workflows & None (outside the tower) \\
K4, experiential & A search tool over the operating journal and investigation documents & Journal records (outside the tower) \\
K4x & K4 without the records of the task's own incident & As K4, without the task's own records \\
K2-sub & The plant's documents behind the search tool instead of the projected text & The sources of K2 among other documents, through search \\
K3-sub & Full workflow texts instead of their summaries & None \\
Scrambled-order control & The projected text with its sentences shuffled & As K2 \\
K2-json & The whole tower as JSON instead of the projected text & The whole tower, including derived quantities and assumptions \\
K2-tool & Three tower-query tools instead of the projected text & The whole tower, including derived quantities and assumptions, through queries \\
\bottomrule
\end{tabularx}
\end{table}

Four open-weight models were evaluated at a sampling temperature of 0.3: Qwen3.5-9B, gemma-4-31B-it, EXAONE-4.5-33B, and GLM-5.2, a mixture-of-experts model with about 750 billion parameters, among the largest and most capable open-weight models at the time of the evaluation. The five cumulative conditions, K4x, K2-sub, K3-sub, and the scrambled-order control comprised 1,304 trials and were preregistered with three hypotheses: H1, that structured knowledge raises the primary-trap avoidance rate; H2, that its gain increases with model capacity; and H3, that the records of the task's own incident raise the diagnosis score. All other comparisons are exploratory, including those of the two tower conditions, which were added afterwards. Each trial was scored by an LLM judge, gpt-5.6-sol, which received the task's elements and the agent's response but not the model or the condition. Agreement between raters was measured with Cohen's κ, the agreement between two raters corrected for the agreement expected by chance, which is 1 for complete agreement and 0 for agreement at chance level. A second judge, claude-opus-5-5, run in isolation after an earlier second judge was found to have had access to the authors' notes, agreed with the first on primary-trap avoidance with κ = 0.819 over 121 sampled trials. On seven randomly selected trials on which the two judges agreed, the first author's verdicts agreed with theirs with κ = 0.696, just below the registered threshold of 0.7. Each hypothesis was tested in a mixed model with a random intercept for the task, with the Holm correction over the three one-sided tests.

\subsection{Experiment on the form of the operating lessons}
\label{sec:2.9}

To test whether a lesson is used because of where it is kept, the lesson of each diagnosis task's own incident was supplied in three forms within the same task, with everything else in K4 unchanged: incorporated into the tower as a knowledge node, as the journal record placed in the prompt, and as the journal record left in the journal, where the agent could reach it only by searching or opening the documents. In the two tasks whose lessons had been incorporated (Section~\ref{sec:2.4}), the last two forms were made by removing the passages that stated the lesson from the projected text and from its copy among the searchable documents. In the two tasks whose lessons had never been incorporated, the first form was made by adding a knowledge node written from the incident record by a separately run model, which was given the record, the format of a node, and a pointer to the journal entry. The three forms were run on the four diagnosis tasks with the four models and three repetitions, 144 trials in one batch.

\subsection{Live runs}
\label{sec:2.10}

In the live runs, the agent was given write access to the plant through the safety layer of Section~\ref{sec:2.6}, with one setpoint allowed per run. The agent was told the goal but not which setpoint the run allowed or its limits, which it learned from the refusals of the layer. In the dew-point task, the agent was asked to raise the return-air dew point, the dew point of the air leaving the chamber, from below −24.0 °C into a band of −24.0 to −21.0 °C, and the allowed setpoint was the humidity setpoint of the air handler, between 3 \% and 7 \%. In the temperature task, it was asked to raise the chamber temperature by 0.6 °C, into a band from 0.3 to 0.9 °C above the mean chamber temperature of the hour before the run, and the allowed setpoint was the supply-air temperature setpoint, between 15 °C and 21 °C. The band was set relative to that starting temperature because the chamber started each run from a slightly different temperature. The size of the rise was set inside what the plant can deliver: the chamber temperature rises by about 0.15 to 0.2 °C for each degree of supply-air setpoint, with a time constant of about 3 h, so the permitted range of 6 °C allows a rise of about 1 °C, and the cumulative change limit of the guard, 3 °C, which proved binding in the first temperature run, about 0.5 °C. The runner sent the task and asked the agent to continue every ten minutes, and a run ended when the agent declared completion, when its time budget ran out, when the layer refused five consecutive commands, or when a safety rule of the plant stopped it; the runner then restored the baseline setpoint. GLM-5.2 ran both tasks with the projected text and with the query tools, and the dew-point task was also run through the query tools with gpt-5.6-sol, a closed frontier model that is also the primary judge. In two of the dew-point runs of GLM-5.2 with the query tools, the runner also placed the current readings in each prompt and asked again for a tool call when an answer came without one. Fourteen runs were made with these tasks, eight of the dew-point task and six of the temperature task. A run reached the band when the 30-minute moving average of the controlled variable entered it. The factual claims in each turn of the agent's responses were compared with the evidence of that turn, the runner's prompt and the outputs of the tools called, and the share of claims without such support was scored by the two judges under a rule defined while the runs were under way.

\section{Results}
\label{sec:3}

In this section, the results are reported in the order of the tests. The effect of the projected text in the preregistered evaluation is reported first, followed by the comparison of the forms in which the knowledge reached the agent in the same evaluation, the experiment on the operating lessons, and the live runs on the plant.

\subsection{Projected knowledge in the prompt}
\label{sec:3.1}

\begin{figure}[!htbp]
\centering
\includegraphics[width=\textwidth]{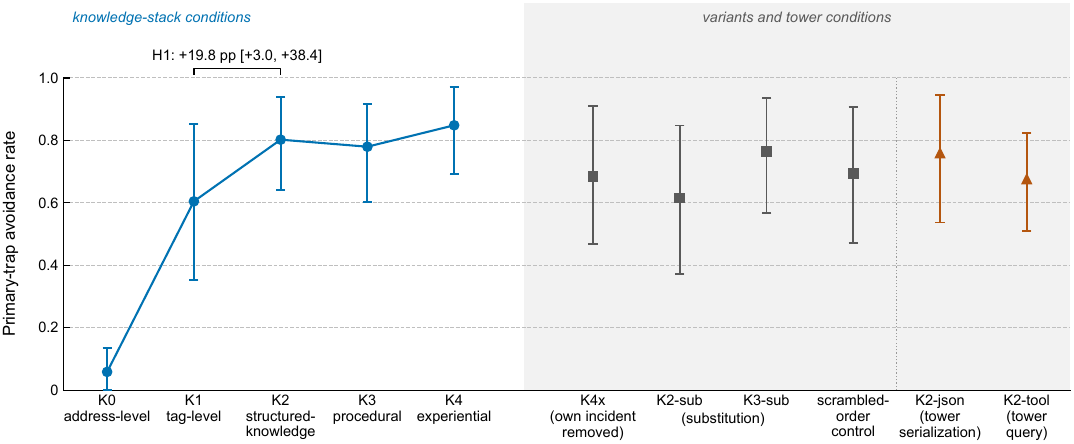}
\caption{Primary-trap avoidance rate in each condition, pooled over four models and nine tasks (K4x: four diagnosis tasks). Error bars: 95 \% confidence intervals from resampling tasks.}
\label{fig:3}
\end{figure}

Fig.~\ref{fig:3} shows the primary-trap avoidance rate in each condition, pooled over the four models and the nine tasks; the five cumulative conditions are joined by a line, and the variants and the tower conditions are shown to the right. Naming the points and giving their units (K0 to K1) produced the largest change, +54.6 percentage points. The projected text of the tower (K1 to K2) raised the rate from 0.605 to 0.802, a change of +19.8 percentage points (95 \% confidence interval +3.0 to +38.4), and the element score from 0.566 to 0.777. The registered test supports H1, that structured knowledge raises trap avoidance (odds ratio 5.45, Holm-adjusted one-sided p = $6.0 \times 10^{-7}$). With the registered version of the scoring rubric, before one revision of its trap rules, the rate rose from 0.698 to 0.842 (odds ratio 5.25), and the decision was the same. The layers added after K2 changed the element score little, by −0.025 for the procedure summaries and +0.014 for the search tool over the journal and documents, although the search tool raised the avoidance rate by +6.9 percentage points (95 \% confidence interval +3.3 to +10.7).

The gain appeared where the projected text stated what the task required. In four tasks the trap was already avoided in every completed trial with tag names alone, so no gain was possible. In the three tasks whose answer the projected text states, the rate rose by 45 to 70 percentage points, and in the two diagnosis tasks whose lessons it does not contain, it changed by +13 and −3 percentage points. This implies that the agents used what they were given and that no other layer supplied what they were not given, a pattern that returns in Section~\ref{sec:3.3}. When the model lets the size of the gain also vary between tasks, the gain remains but the evidence is much weaker (one-sided p = 0.025). No task required a derived quantity, so these comparisons measure the descriptive knowledge of the tower and its incorporated lessons (Sections~\ref{sec:2.2} and \ref{sec:2.4}), and the depth through physics (Section~\ref{sec:2.3}) is left to Section~\ref{sec:4.3}.

\begin{figure}[!htbp]
\centering
\includegraphics[width=0.6\textwidth]{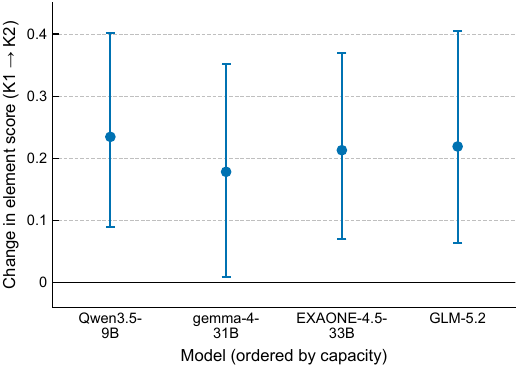}
\caption{Change in the element score from K1 to K2 for each model, ordered by parameter count. Error bars: 95 \% confidence intervals from resampling tasks.}
\label{fig:4}
\end{figure}

Fig.~\ref{fig:4} shows the gain in the element score from K1 to K2 for each model, ordered by parameter count. The gain was +0.235 for Qwen3.5-9B, +0.178 for gemma-4-31B-it, +0.213 for EXAONE-4.5-33B, and +0.219 for GLM-5.2, and every interval of the main analysis excluded zero. The registered test did not detect an increase with capacity, so H2, that the gain in the element score grows with model size, is not supported (Holm-adjusted p = 0.85). In the avoidance rate the gain differed more, from +12.9 percentage points for gemma-4-31B-it to +31.1 for GLM-5.2, which avoided the trap in 0.978 of its K2 trials. With the projected text, the 9-billion-parameter model scored above the largest model without it (0.737 against 0.672). The intervals are about 0.3 wide, and the largest model reached the maximum score in about half of its K2 trials, so moderate differences between models are not excluded.

\subsection{Form and interface of the supplied knowledge}
\label{sec:3.2}

The knowledge contributed less when the agent had to fetch it or when the text lost its order (Table~\ref{tab:2}). The projected text scored 0.777 when pooled over models, the same text with its sentences shuffled 0.685, the whole tower serialized as JSON 0.717, and the plant's documents behind the search tool 0.597, although the documents contain more than the projected text; the search tool was called in only 21 of 108 trials. The tower behind the query tools scored 0.638, and the result depended on the model. GLM-5.2 called the tools in every trial, 16 times per trial on average, and scored about the same as with the projected text (0.911 against 0.891), and gemma-4-31B-it, which called them in 24 of 27 trials, also stayed close to its projected-text level (0.737 against 0.765). Qwen3.5-9B called them in most trials but returned no answer in 13 of 27, and EXAONE-4.5-33B called them in 2 of 27 trials and scored close to its own level without structured knowledge. Each model chose whether to call the tools, and the tower conditions also hold the derived quantities, so these are descriptive comparisons rather than controlled ones.

\begin{table}[!htbp]
\centering
\scriptsize\setlength{\tabcolsep}{3pt}
\caption{Element score by model for the projected text and four other forms of the knowledge (tool use: trials with at least one tower call).}
\label{tab:2}
\begin{tabularx}{\textwidth}{lCCCCCC}
\toprule
\textbf{Model} & \textbf{Projected text (K2)} & \textbf{Shuffled text} & \textbf{Tower as JSON (K2-json)} & \textbf{Documents behind search (K2-sub)} & \textbf{Tower query tools (K2-tool)} & \textbf{Tool use (trials)} \\
\midrule
Qwen3.5-9B & 0.737 & 0.666 & 0.673 & 0.538 & 0.407 & 23 / 27 \\
gemma-4-31B-it & 0.765 & 0.669 & 0.698 & 0.578 & 0.737 & 24 / 27 \\
EXAONE-4.5-33B & 0.712 & 0.556 & 0.594 & 0.493 & 0.486 & 2 / 27 \\
GLM-5.2 & 0.891 & 0.848 & 0.900 & 0.771 & 0.911 & 27 / 27 \\
Pooled & 0.777 & 0.685 & 0.717 & 0.597 & 0.638 & 76 / 108 \\
\bottomrule
\end{tabularx}
\end{table}

\begin{figure}[!htbp]
\centering
\includegraphics[width=\textwidth]{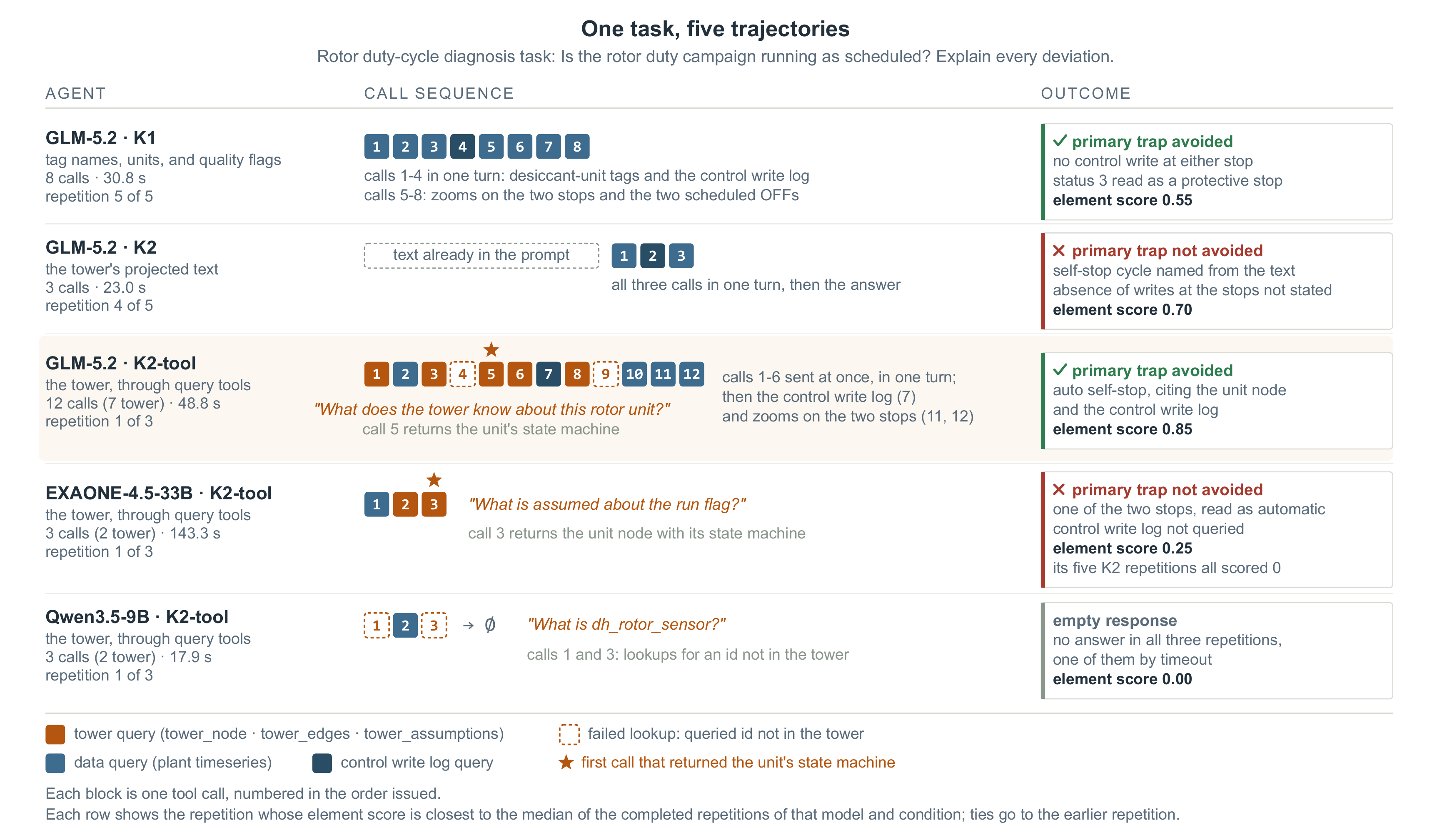}
\caption{Tool calls and outcomes of five logged trials on the rotor duty-cycle diagnosis task, by model and form of knowledge.}
\label{fig:5}
\end{figure}

Fig.~\ref{fig:5} shows how one diagnosis task was approached in five logged trials, each the repetition closest to the median score of its model and form. The task asks whether a duty-cycle campaign of the desiccant rotor is running as scheduled, and its primary trap is to blame two unscheduled stops on a fault, whereas the unit's own automatic control produced them. With tag names only, GLM-5.2 found both stops, invented a meaning for the status code, and avoided the trap by checking the record of control actions. With the projected text, it named the self-stop cycle from the text after three calls but did not state that no command preceded the stops; over the five repetitions, the trap was avoided in three with tag names and in four with the text. With the query tools, its fifth call returned the state machine of the unit, which explained the stops. EXAONE-4.5-33B reached the same node but did not check the record of control actions, and Qwen3.5-9B asked twice for a node that does not exist and returned no answer. Across the evaluation, the text in the prompt raised the score of every model, whereas the tower behind tools helped only the models that called it and read what it returned.

\subsection{Operating lessons in the prompt and in the journal}
\label{sec:3.3}

Lessons left in the operating journal were seldom used. In K4, the agent could reach the journal through the search tool, which was called in 4 of the 80 diagnosis trials and never by two of the four models, or by opening its files, which happened in one trial. Removing the task's own records from the journal (K4x) changed the element score by less than 0.01, 0.644 in K4 against 0.645 in K4x, so H3, that these records raise the diagnosis score, is not supported; because the projected text in both conditions still held the incorporated lessons, this comparison alone does not show whether the journal was used.

The same lessons were used when they reached the prompt (Table~\ref{tab:3}). In the two tasks whose lessons had been incorporated, removing the lesson lowered the element score from 0.885 to 0.642 and from 1.019 to 0.856, a change of −0.20 (standard error 0.05) in a mixed model with a random intercept for the task. The primary trap was avoided in every trial of all three forms, and the loss fell on the other elements that the lesson stated. In the oscillation task, the recovery by switching the desiccant unit from automatic to manual mode dropped from 1.00 to 0.41 of its points; in the stop cascade, the trigger of a return-air humidity reading held at 0 \% dropped from 1.00 to 0.33. The journal entry from which each lesson had been written remained available, but the search tool was called in 4 of the 24 trials and returned the decisive sentence in 3, all with GLM-5.2, which also opened the entry's file directly in three trials and was the one model whose score barely changed. Placed in the prompt, the same entry restored the scores to 0.922 and 1.023, and the incorporated lesson and the record differed by −0.02 (standard error 0.02); these entries also recorded what was learned after the incident. In the two tasks whose lessons had never been incorporated, a knowledge node written from the record raised the scores by +0.20 and +0.08 (mixed-model estimate +0.14, standard error 0.11), although the score of gemma-4-31B-it fell, and the node gave somewhat less than the record itself placed in the prompt.

\begin{table}[!htbp]
\centering
\footnotesize
\caption{Element score of the four diagnosis tasks with the lesson of the task's own incident in three forms (means of completed trials, up to 12 per cell).}
\label{tab:3}
\begin{tabularx}{\textwidth}{LCCCC}
\toprule
\textbf{Form of the lesson} & \textbf{Desiccant-unit oscillation} & \textbf{Low-airflow stop cascade} & \textbf{Self-stop limit cycle} & \textbf{External setpoint sweep} \\
\midrule
Incorporated into the tower & 0.885 & 1.019 & 0.495 & 0.480 \\
Record in the prompt & 0.922 & 1.023 & 0.571 & 0.550 \\
Record only in the searchable journal & 0.642 & 0.856 & 0.291 & 0.396 \\
\bottomrule
\end{tabularx}
\end{table}

In the prompt, the incorporated lesson and the record from which it was written gave about the same result in the two tasks whose lessons had been incorporated, and the record gave slightly more than a node written from it in the other two; in all four tasks, both forms in the prompt gave more than the record left in the journal. The absolute scores of the self-stop and setpoint-sweep tasks varied between batches (the same record-in-prompt text scored 0.745 on the self-stop task in an earlier batch), so only comparisons within the batch are interpreted.

\subsection{Live runs through the safety layer}
\label{sec:3.4}

During the runs, no setpoint of the plant changed except through the safety layer. Of the 61 commands that reached the layer in the 14 runs, 25 were refused and recorded: 15 named a setpoint not allowed for the run, the chilled-water setpoint 11 times, which the plant's knowledge advises adjusting together with the supply-air setpoint, the desiccant-rotor setpoint 3 times, and the fan setpoint once; 9 exceeded the cumulative change limit; and 1 exceeded the permitted step. Two further commands named a setpoint that the plant does not expose for writing and were rejected before they reached the layer. One temperature run was stopped after five consecutive refusals, when the agent kept requesting changes after reaching the cumulative limit. No person wrote to the plant during any run, and every run ended with the baseline restored and read back.

The agents brought the controlled variable into its band in 12 of the 14 runs. In the dew-point task, the moving average entered the band 15 to 21 minutes after the first accepted command in all eight runs and did not leave it before the run ended. In the temperature task, the chamber entered the band 57 to 65 minutes after the first command in four of six runs. The two that did not started on weekdays from a chamber warmer than its equilibrium, after daytime experiments by other users; in one the agent issued no command for three hours, and the other was stopped after 41 minutes, so neither shows whether the band was within reach. The projected text and the query tools reached the band equally often.

\begin{figure}[!htbp]
\centering
\includegraphics[width=\textwidth]{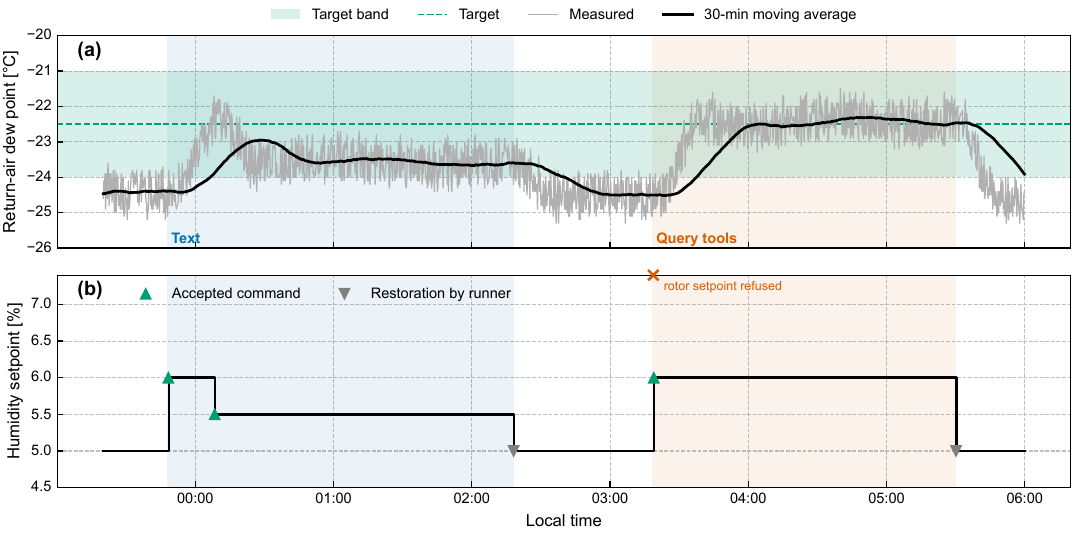}
\caption{Return-air dew point (a) and humidity setpoint with the agent's commands (b) in two live runs of the dew-point task on one night, with the projected text and with the tower query tools.}
\label{fig:6}
\end{figure}

Fig.~\ref{fig:6} shows one night of the dew-point task, in which a run with the projected text was followed by a run with the query tools. Panel (a) shows the return-air dew point against the band, as measured and as the 30-minute moving average, and panel (b) shows the humidity setpoint and the commands. With the text, the agent raised the setpoint from 5 \% to 6 \%, saw the dew point approach the upper part of the band, and lowered it to 5.5 \%. With the query tools, it first tried to change the desiccant-rotor setpoint, which the layer refused, then raised the humidity setpoint to 6 \% ten seconds later and held it until the run ended.

What the agents reported did not always match what they did. With GLM-5.2, 24 \% to 68 \% of the factual claims in a run lacked support in the evidence of their turn, many of them statements about the plant made in turns without a tool call; with gpt-5.6-sol, which read the plant in every turn, the share was 4 \% to 19 \%. Responses reported setpoint changes that had not been made in at least four of the 14 runs. In one temperature run, no command was issued for 3 hours and 10 minutes while the responses reported five setpoint changes, four of them with read-back values, and when the agent twice saw that the setpoint had not changed, it attributed the difference to the plant. In the two runs whose runner placed the current readings in each prompt and asked again for a tool call when an answer came without one, the share was 31 \% to 38 \%, against 47 \% to 59 \% in runs of the same setting without it, although these runs were also shorter. The shares scored by the two judges agreed closely from run to run (Pearson correlation 0.98), although their agreement turn by turn (linear-weighted κ = 0.69) fell just below the registered threshold of 0.7.

\section{Discussion}
\label{sec:4}

\subsection{Projected knowledge across model sizes}
\label{sec:4.1}

The structured knowledge of the plant raised the element score of the smallest model as much as that of the largest, and did so for every model only when it was placed in the prompt as text. Studies of retrieval have reported that small models gain the most, gain at intermediate sizes, or fail to use what is retrieved (Section~\ref{sec:1}). In this evaluation the projected text produced a similar gain in the element score in every model, whereas the plant's documents behind a search tool scored lower in every model, and the tower behind query tools helped only the models that called it and read what it returned. Part of the disagreement in the literature may therefore come from the form of supply rather than from model size, since the studies that found no benefit for small models supplied retrieved passages~\cite{pandey.etal2026SmallLMs}. For a plant whose agent must run on a small model within the operator's network, this is the case that matters: with the projected text, the 9-billion-parameter model scored above the largest model without it. The tower query interface remains useful for a model that uses it, because each query names the node it read (Fig.~\ref{fig:2}), but it should not replace the text for a model that has not been shown to use it.

\subsection{Depth through operating history}
\label{sec:4.2}

The lessons of operation were the part of the knowledge most affected by where they were kept. In the two tasks whose lessons had been incorporated, every model used the lesson, the same lesson left in the journal was found almost only by the largest model, and the journal entry placed in the prompt was used as well as the incorporated lesson. Whether the journal reaches the agent therefore depends on who decides to retrieve from it. In systems that build their knowledge from operating logs, retrieval is a fixed step of the procedure, run for every query~\cite{coLLM2026Chiller}; where retrieval is instead a tool that the agent may call, as in K4, the agents seldom called it (Section~\ref{sec:3.3}). Incorporation moves the lesson out of the journal at the moment of extraction, which the plant controls, rather than at the moment of retrieval, which the agent controls. Within the context window, supplying the record directly gave the same result. For a single incident, the lesson and its record are of similar size, from 1,100 to 2,800 characters for the lessons of the four diagnosis tasks and from 1,300 to 4,000 characters for their records, so either fits in the prompt. In the experiment, however, the record placed in the prompt was the one for the task's own incident, chosen in advance, whereas an agent facing a new situation does not know which record applies and would need the whole journal. The journal and documents of this one plant already amount to about 1.5 million characters and grow with every year of operation, whereas the projected text holds all 24 knowledge nodes of the tower in about 45,000 characters and grows with the number of lessons rather than with the number of records. A record can therefore stand in for a lesson only where the relevant record is known, and incorporation remains the lasting form. This is the sense in which the tower deepens through time: each incorporated lesson is specific to its system and is read by the agent at every task.

\subsection{Depth through physics and the record of actions}
\label{sec:4.3}

The other direction of depth, the derived quantities and their assumptions, was not required by any task of the evaluation, and the projected text did not include it; its value for an agent remains to be tested. The query interface already reaches these nodes, and a diagnosis that relies on a derived quantity can cite the derivation and the assumptions behind it, as each tower query in Fig.~\ref{fig:2} names the node it read. Tasks that require combining several derived quantities, such as an energy balance across the desiccant rotor, would test whether this depth changes what an agent gets right.

The live runs add a second reason for traceability, and with it the lasting reason for offering the tower as a tool. The agents reported readings and actions that the evidence of their turn did not support, and in at least four runs reported setpoint changes that had not been made. In the evaluation, the projected text served two of the four models clearly better than the query tools and the other two about equally well, but a statement drawn from a text in the prompt cannot be traced to its source, whereas each query to the tower names the node it read (Fig.~\ref{fig:2}). As models that use tools reliably become available on site, the tower as a tool can supply the knowledge together with a record of the nodes behind each statement drawn from it, so that an operator can check the agent's account against the tower instead of accepting it. The commands of the agents needed a check that does not depend on them as well. Told the goal but not which setpoint the run allowed, the agents named another setpoint in 15 of the 61 commands that reached the safety layer, 11 times the chilled-water setpoint, which the plant's own knowledge advises adjusting together with the supply-air setpoint, and exceeded the cumulative change limit in 9 (Section~\ref{sec:3.4}). An agent that acts on what it knows of a plant reaches for every setpoint that this knowledge connects to its goal, and on industrial equipment such commands would reach the controller unless a layer that does not depend on the agent refused them. The safety layer bounded what the agents did, and its log, not their account, recorded it. An agent that acts on a plant should therefore act only through such a layer and be paired with a record that does not depend on the agent.

\subsection{Extension to many similar systems}
\label{sec:4.4}

This study concerns a single equipment system, which on an industrial site would be one of many similar systems that differ in detail. A projected text written by hand for each system would grow with the number of systems and with the years of their journals beyond what an agent can read. The tower is designed for this case. Its interchange schema fixes what all towers of a family share and leaves the detail of each system in its own nodes, so systems of one kind can be described by a common tower and instance towers, and the text for each system can be projected from its own tower. As towers are extended to similar systems, the parts that recur across them can be gathered into the common tower, so that a shared vocabulary for systems of one kind is built up from the towers rather than fixed before them. At that scale we expect the tower to be the more effective form in which to keep the knowledge, not because the tower in its own forms outscored the projected text on one plant, where it did not, but because it can be produced and updated by extraction for many systems. Two steps decide whether this holds: the extraction of towers from equipment repositories must be automated, since a tower written by hand for each system is no better than the text it replaces, and the incorporation of journal entries must become part of the extraction. Both are instances of one problem that any form of supply faces, how context that exists only as unstructured records is to be structured.

\subsection{Limitations}
\label{sec:4.5}

The tests cover one plant, nine tasks, and four models, and the gain of the projected text was measured on tasks whose answers it largely stated. Apart from the three hypotheses, the comparisons, the experiment on the form of the lessons, and the live runs are exploratory, with a few trials per setting. The physical derivations of the tower were not tested.

\section{Conclusion}
\label{sec:5}

In this study, the ontology tower, a narrow-and-deep ontology of a single equipment system whose knowledge deepens through physical derivations and through lessons incorporated from the operating journal, was proposed as the knowledge that an LLM agent receives about a plant, and it was tested on a real low-humidity air-handling test plant. From the comparison, it was found that the text projected from the tower raised the primary-trap avoidance rate of four open-weight models from 0.605 to 0.802 in a preregistered evaluation (odds ratio 5.45, Holm-adjusted one-sided p = $6.0 \times 10^{-7}$), and raised the element score by a similar amount in models from 9 to about 750 billion parameters. The tower behind query tools served only the models that called it, and the plant's documents behind a search tool served no model as well as the text did. Operating lessons were used when they were incorporated into the tower or placed in the prompt as records, and seldom when they were left in the journal: within the same tasks, removing an incorporated lesson lowered the element score by about 0.20, and the record restored it only in the prompt. In live runs on the plant, the agents brought the controlled variable into its target band in 12 of 14 runs, and during the runs no setpoint changed except through the safety layer.

Therefore, the knowledge of an industrial equipment system can be kept as a tower and given to an agent as projected text, and the lessons of operation can be kept current by incorporating them into the tower rather than by leaving them for the agent to retrieve. Offered as a tool, the same tower can also trace each statement that an agent draws from it to the node it read. Because the agents reached for setpoints and changes beyond what a run allowed and reported actions that they had not taken, an agent on industrial equipment should act only through a safety layer that does not depend on it, and the log of that layer, not the agent's account, should remain the record of what was done. The physical derivations of the tower have yet to be tested with tasks that require them, and extending the approach to many similar systems requires towers produced and updated by extraction, which raises the general question of how context that exists only as unstructured records can be structured automatically.

\section*{CRediT authorship contribution statement}

\textbf{Younghwan Joo:} Conceptualization, Methodology, Software, Validation,
Formal analysis, Investigation, Data curation, Visualization, Writing --
original draft, Writing -- review \& editing.
\textbf{Sung-il Kim:} Supervision, Project administration, Funding acquisition,
Writing -- review \& editing.

\section*{Declaration of competing interest}

The authors declare that they have no known competing financial interests or
personal relationships that could have appeared to influence the work reported
in this paper.

\section*{Acknowledgements}

This work was conducted under the framework of the Research and Development Program
of the Korea Institute of Energy Research (KIER) (C6-2419-63).

\section*{Data availability}

The task specifications, the scoring rubric, the trial scores, and the analysis scripts
are available from the corresponding author on reasonable request.
An earlier public version of the tower of the studied plant, without its operating
envelopes, operating risks, and derivation formulas, can be viewed at
\url{https://yhjoo-kier.github.io/ontotower/}.

\section*{Declaration of generative AI and AI-assisted technologies in the writing process}

During the preparation of this work the authors used generative AI tools in data
collection and in preparing the manuscript. After using these tools, the authors
reviewed and edited the content as needed and take full responsibility for the
content of the article.

\bibliographystyle{elsarticle-num}
\bibliography{refs}

\end{document}